\documentclass[12pt,pra,aps,amssymb,amsmath,tightenlines]{revtex4}

\usepackage{dsfont}
\usepackage{amsmath, amsthm}
\usepackage{amssymb, graphicx}
\newcommand{\half}{\mbox{$\textstyle \frac{1}{2}$}}

\newcommand{\re}{\mbox{$\rm e$}}

\newcommand{\rd}{\mbox{$\rm d$}}

\begin{document}

\title{Theory of Information Pricing}

\author{Dorje~C.~Brody${}^1$ and Yan Tai Law${}^2$}

\affiliation{${}^1$ Mathematical Sciences, Brunel University, Uxbridge UB8 3PH, UK\\ 
${}^2$ Department of Mathematics, Imperial College London, London SW7 2AZ, UK}

\date{\today}

\begin{abstract}
In financial markets valuable information is rarely circulated homogeneously, because of time required for information to spread. However, advances in communication technology means that the `lifetime' of important information is typically short. Hence, viewed as a tradable asset, information shares the characteristics of a perishable commodity: while it can be stored and transmitted freely, its worth diminishes rapidly in time. In view of recent developments where internet search engines and other information providers are offering information to financial institutions, the problem of pricing information is becoming increasingly important. With this in mind, a new formulation of utility-indifference argument is introduced and used as a basis for pricing information. Specifically, we regard information as a quantity that converts a prior distribution into a posterior distribution. The amount of information can then be quantified by relative entropy. The key to our utility indifference argument is to equate the maximised \textit{a posterior} utility, after paying certain cost for the information, with the \textit{a posterior} expectation of the utility based on the \textit{a priori} optimal strategy. This formulation leads to one price for a given quantity of upside information; and another price for a given quantity of downside information. The ideas are illustrated by means of simple examples. 
\end{abstract}
 
%\pacs{no pacs}

\maketitle

%\newpage

%\section{Introduction}

The role and importance of information and its costing has long been recognised 
in economics literature \cite{Hayek}. Nowadays, people often obtain information 
through internet search engines; but the very concept that one should search price 
information and utilise it to reduce expenditure, in the context of commodity 
markets, has been noted for at least half a century \cite{Stigler}, if not considerably 
longer. Nevertheless, the use of information has never been as important in 
the past. Indeed, accesses to internet search engines---whose objectives are in 
information provision---constitute an integrated part of the lives of millions. 
Internet retailers likewise not only sell products but also provide a range of 
information via their recommendation engines, helping consumers find what they 
like but along the way also yielding additional profits. The quality of a 
recommendation engine, however, is difficult to assess if one cannot assign 
values to the information it provides. 

Rapid developments in communication technology have their upsides as well as 
downsides. On the one hand, access to targeted information has been made easy 
(for example, when one knows the relevant {\scriptsize URL} from which information 
can be obtained), while on the other hand a generic search for specific information 
is becoming harder due to the flooding of irrelevant information that constitute 
noise. This is a phenomenon envisaged by Wiener, who used the second law 
of thermodynamics to explain the fact that in the long run one cannot overcome 
the impact of noise that overwhelms useful information, and that this effect is 
in part due to the advancements of communication technology \cite{wiener}. 
Wiener also recognised that this inevitable flow can locally be reversed by 
means of innovations. It is for this reason that internet search engines and other 
information providers are constantly on alert of innovative ideas for information 
extraction and provision. Evidently, enhancement of information extraction and 
provision, and implementation of innovation, cannot be realised without cost. As 
a consequence, access to valuable information will likewise entail some costs. 
The question then is: How do we price information? 

There is a form of information that is regarded as particularly valuable---the kind 
of information that allows one to anticipate, at least with some statistical 
significance, what is to come, that is, to anticipate future trends. For example, 
suppose that a company has just released a new product. A search engine can 
access various internet sites and analyse texts (e.g., blog articles and tweets) 
to determine how satisfied  consumers are with the product. If customer 
satisfaction is high, the search engine can anticipate that future sales will 
increase and hence the company will perform well. Of course, neither the data 
source nor the analysis can be perfect; nevertheless, if the methodology of 
information extraction is adequate, then the prediction of the search engine can 
be statistically significant. After a while, however, sales figures are released, by 
which time the worth of this information diminishes. 

The above example illustrates the idea of information diffusion. The search 
engine extracts information from in principle publicly available sources. Hence 
one need not be a large search engine to obtain the same information at the 
same time. However, in practice most people do not have the resource to 
analyse in real time millions of articles available on the internet. Nevertheless, 
after sufficient amount of time (which can be short or long), important information 
will be appreciated by a broad mass (and thus reflected in the share price, in the 
above example), diminishing the worth of information as providing predictive 
insights. At this stage, information has truly reached the 
public domain, but this is necessarily some time after its initial appearance 
somewhere in `public'. 

From the viewpoint of a search or recommendation engine, and given their 
resources, capabilities, and objectives, therefore, it makes sense to extract 
useful information before the public as a whole has had chance to do so, and 
sell it off to a third party who is in the position to utilise this information. We 
can thus think of information as a tradable asset, albeit that it will have an 
expiry date. In order to price information, however, we must cast 
these intuitive ideas in terms of simple but precise mathematical language. 
The purpose of the present paper is to show how this is done. 
Our analysis reveals the surprising fact that the value of information in general 
need not be monotonic in the quantity of information, because sometimes less 
information is more useful than more information. 

To illustrate how information can be valued, let us consider a simple example. 
Suppose that we 
are interested in the price tomorrow of a stock, whose current price is $s_0$. Our 
formulation of information pricing is applicable to a broader class of situations 
involving anticipating uncertain future events. Nevertheless, we find that the 
example concerning information for predicting future market very convenient for 
providing intuitive understanding of the theory (cf. \cite{BBMK}). To 
keep the discussion simple, let us assume a single-period setup so that the price 
$s_0$ today will become price $s_1$ tomorrow. The value of $s_1$ is of course 
unknown today, but we assume that it satisfies a known probability law $p(x)$. 
That is, we regard $s_1$ as a random variable defined on a probability space 
$(\Omega,{\mathcal F},{\mathbb P})$, where ${\mathbb P}$ denotes the physical 
probability measure. A search engine now performs text mining to extract 
information relevant to determining the value of $s_1$. Evidently, 
no search engine is capable of identifying the value of $s_1$; at best, it gathers 
noisy information in the form of \textit{signal plus 
noise}. In the simplest setup, $s_1$ itself constitutes the signal, but a search 
engine can only extract the value of 
\begin{eqnarray}
\xi = s_1 + \epsilon, \label{eq:1}
\end{eqnarray}
where $\epsilon$ represents noise, independent of $s_1$, with known density 
$f(x)$. In other words, a search engine can sample the value of $\xi$, which 
contains new information that is not already encoded in the value today of the 
stock, but not sufficient to determine the value tomorrow of the stock. With the 
knowledge of $\xi$ the prior density $p(x)$ of $s_1$ can thus be updated to the 
posterior density 
\begin{eqnarray}
\pi(x|\xi) = \frac{p(x) f(\xi-x)}{\int \rd x\, p(x) f(\xi-x)}
\label{eq:2}
\end{eqnarray}
in accordance with the Bayes formula. Writing $\pi_\xi(x)=\pi(x|\xi)$ for the 
posterior density we define the concept of information provision as follows: 
It is an operation that supplies the updated probability law:
\begin{eqnarray}
p(x) \to \pi_\xi(x). \label{eq:3} 
\end{eqnarray}
Stated more precisely, information provision is an updating of probability law in 
the form of (\ref{eq:3}) where $\pi_\xi(x)$ is necessarily adapted to a larger 
filtration (i.e. information set or knowledge) than that of $p(x)$. 

Having defined the notion of information provision, we proceed to consider the 
pricing issues. Let us first illustrate the idea, and then try to formalise it. In the 
above example, an uninformed investor will use the knowledge $p(x)$ to 
determine the optimal asset allocation, which may contain certain amount of 
stock $s_0$, given the initial budget $W_0$. An informed investor is one who 
has purchased the knowledge $\pi_\xi(x)$ at a cost $c$ and used this knowledge 
to determine the optimal asset allocation, given a smaller initial budget $W_0-c$. 
For example, if the $p$-likelihood of stock price going up is high while the 
$\pi$-likelihood of this event is low, then the informed investor will purchase 
smaller amount of stock so as to circumvent likely loss. The problem then is 
to identify a fair value of the cost $c$. We would like to use a version of utility 
indifference argument to determine the cost. However, as we shall indicate below, 
for information pricing, the standard utility argument has to be augmented in a 
subtle manner so as to determine the fair price. Let us illustrate this by 
means of a simple example. 

\noindent \textit{Example 1.1: uninformed investor}. Consider a portfolio consisting 
of a single risky asset $s_0$ and a risk-free money market (bank) account. Let 
$\theta$ be the amount allocated to the money market account, and $\phi$ be the 
unit of stock invested. Then starting with the initial wealth $W_0$ we have the 
allocation 
\begin{eqnarray}
W_0 = \theta + \phi s_0. \label{eq:4}
\end{eqnarray}
Let $\delta$ be the discount factor over the time period so that a unit 
cash invested in the money market account will yield $\delta^{-1}$ at the end of 
the period. Then the terminal wealth becomes 
\begin{eqnarray}
W_1 = \theta \delta^{-1} + \phi s_1.
\end{eqnarray}
The uninformed investor will determine the optimal asset allocation strategy 
$(\theta^*,\phi^*)$ such that the expected utility ${\mathbb E}_p[U(W_1)]$ of 
the terminal wealth is maximised. So for example if $s_1$ is a binary random 
variable taking values $\{0,1\}$ with probabilities $\{p, 1-p\}$, and if the investor 
has an exponential utility 
\begin{eqnarray}
U(x) = 1-\exp(-\alpha x) \qquad \alpha \in \mathbb{R}^+,
\end{eqnarray}
then we have 
\begin{eqnarray}
{\mathbb E}_p[U(W_1)] = 
1- \re^{-\alpha \delta^{-1} (W_0-\phi s_0)} \left( p + (1-p) \re^{-\alpha \phi} \right),
\end{eqnarray}
where we have eliminated $\theta$ by use of (\ref{eq:4}). Maximising this over 
$\phi$ and making use of (\ref{eq:4}), and writing ${\bar s}_0=s_0 \delta^{-1}$ for 
the forwarded stock price, we obtain the following optimal asset allocation 
strategies:
\begin{eqnarray}
\phi^* = \frac{1}{\alpha}\ln \left[ \frac{(1-p)(1-{\bar s}_0)}{p {\bar s}_0} \right], 
\qquad \theta^* = W_0-\phi^* s_0. \label{eq:8} 
\end{eqnarray}
The optimal expected utility of terminal wealth is thus 
\begin{eqnarray}
u(W_0) = 1- \re^{-\alpha \delta^{-1} (W_0-\phi^* s_0)} \left( p + (1-p) 
\re^{-\alpha \phi^*} \right) . 
\end{eqnarray} 
Note that the optimisation has been performed without constraints. If short selling 
of either asset is forbidden, $(\theta^*, \phi^*)$ are either $(0,W_0/s_0)$ or 
$(W_0,0)$ depending on whether $\phi^*\geq W_0/s_0$ or $\phi^*<0$; otherwise, 
$(\theta^*,\phi^*)$ are given by (\ref{eq:8}). \hfill$\diamond$

We now consider the action of an informed investor in this simple example. 
We assume that there is an information provider who has the ability to sample 
\begin{eqnarray} 
\xi = s_1 + \epsilon,
\end{eqnarray} 
where $\epsilon$ is a noise term independent of $s_1$, which for simplicity of 
exposition we assume to take values $\{0,1\}$ with probabilities $\{q, 1-q\}$. In 
this example the value of $\xi$ can be $0$, $1$, or $2$, and the \textit{a 
posteriori} probability $\pi_\xi$ that $s_1=0$ is given by 
\begin{eqnarray} 
\pi_0 &=& 1 \nonumber \\
\pi_1 &=& \frac{p(1-q)}{p(1-q)+q(1-p)}\\
\pi_2 &=& 0. \nonumber 
\end{eqnarray}
Since the information provider has to have invested capital in developing a system 
that is capable of sampling $\xi$, and since it has to continue investments for 
system maintenance, it is only reasonable to expect a payment for the provision of 
this information. Before we proceed to value this cost, let us first examine the 
strategy of an informed investor. 

\noindent \textit{Example 1.2: informed investor}. Let $c$ be the cost for the 
purchase of information. Then the initial budget condition for the informed 
investor becomes 
\begin{eqnarray} 
W_0 = \vartheta + \varphi s_0 +c, \label{eq:13}
\end{eqnarray} 
where $(\vartheta,\varphi)$ denote portfolio positions of the informed investor. 
An application of the optimisation procedure described above then yields: 
\begin{eqnarray}
\varphi^* = \frac{1}{\alpha}\ln \left[ \frac{(1-\pi_\xi)(1-{\bar s}_0)}{\pi_\xi {\bar s}_0} 
\right], \qquad \vartheta^* = W_0-c-\varphi^* s_0 ,  \label{eq:14}
\end{eqnarray} 
where $\pi_\xi = {\mathbb P}(s_1=0|\xi)$. In this case, utility of the optimal terminal 
wealth is given by
\begin{eqnarray}
u_c(W_0) = 1- \re^{-\alpha \delta^{-1} (W_0-c-\varphi^* s_0)} 
\left( \pi_\xi + (1-\pi_\xi)\re^{-\alpha \varphi^*} \right). \label{eq:15}
\end{eqnarray} 
Similarly to the previous example, these results are based on unconstrained 
optimisation. If short selling is not permitted, for instance, then these results 
hold if $0 \leq \varphi^* \leq (W_0-c)/s_0$; otherwise we must take appropriate 
boundary values. \hfill$\diamond$

We are now in the position to discuss valuation of information in the above 
examples. To this end we would like to employ an argument based on utility 
indifference, similar to the one presented, for example, by Grossman and Stiglitz 
\cite{GrossmanStiglitz}. The argument intuitively goes as follows. If possession of 
extra information at no cost provides on average a better payoff, then every investor 
would sought to be informed. Consequently, there should be a positive value 
assigned to this information. If on the other hand the price is 
too high, then no one would want to be informed. Hence there exists an equilibrium 
level for the cost of information at which investors are indifferent, as far as their 
preferences are concerned. 

Naively, one might then identify the utility indifferent price of information by solving 
the following equation for $c$: 
\begin{eqnarray}
u(W_0) = u_c(W_0) \label{eq:16}
\end{eqnarray}
so that the expected utility of the \textit{a posteriori} optimal strategy after 
paying for the information equals the utility of the \textit{a priori} optimal strategy. 
This is in effect the pricing formula proposed by Amendinger \textit{et al}. 
\cite{Schweizer} (see also \cite{Imkeller}). 
The intuition behind this relation is as follows: Provided that the cost $c$ is 
sufficiently small, the possession of additional relevant knowledge, in the form of 
a strictly larger filtration, necessarily increases expected utility. Since utility is a 
decreasing function of the cost $c$, there must be a positive number $c^*$ such 
that (\ref{eq:16}) is satisfied for $c=c^*$. 
In the context of information pricing, however, this intuitive argument fails to 
determine the correct value of $c^*$. With a further reflection, the reason for this 
becomes apparent: Ignorance can make people happy. While it is true that with 
the additional knowledge an investor will on average perform better (cf. \cite{BDFH}), 
an uninformed 
investor can be more optimistic about the future outlook. Take, for instance, an 
extreme case where the \textit{a priori} probability that the stock price moving up 
by a significant amount is high. An uninformed investor will thus put the majority 
of the initial wealth into purchasing this stock. Suppose, however, that the 
\textit{a posteriori} probability indicates that the stock price is more likely to move 
down. An informed investor will thus only invest a small fraction of the initial 
wealth into this stock to avoid a large loss. In this case, a likely scenario is that 
an informed investor knows that there is little prospect of making a large profit (i.e. 
small utility), but manages to prevent a loss, while an uninformed investor has a 
very high hope (i.e. large utility), but ending up with a loss. According to the pricing 
formula (\ref{eq:16}), therefore, the information cost becomes negative in such a 
scenario. 

Grossman and Stiglitz \cite{GrossmanStiglitz} (see also \cite{ver,kyle,peress}), 
on the other hand, propose the use of the following identity to fix the cost: 
\begin{eqnarray}
{\mathbb E}_p[U(W_p^*)] = {\mathbb E}_p[U(W_{\pi,c}^*)]. \label{eq:17.0} 
\end{eqnarray}
Here, $W_p^*=\theta^* \delta^{-1} + \phi^* s_1$ denotes the random variable 
associated with the terminal wealth based on the implementation of the 
\textit{a priori} optimal strategy (\ref{eq:8}); whereas $W_{\pi,c}^*=\vartheta^* 
\delta^{-1} + \varphi^* s_1$ denotes the random variable associated with the 
terminal wealth based on the implementation of the \textit{a posteriori} 
optimal strategy (\ref{eq:14}), having paid $c$ for the information. In the 
present example, (\ref{eq:17.0}) leads to the pricing formula 
\begin{eqnarray}
c^* = \frac{\delta}{\alpha}\ln \left[ \frac{1}{q^{{\bar s}_0}(1-q)^{1-{\bar s}_0}} \right] . 
\label{eq:17.1} 
\end{eqnarray}
The shortcoming of the Grossman-Stiglitz pricing formula (\ref{eq:17.0}), however, 
is that the cost is fixed irrespective of its quality. For example, if the realised 
value of $\xi$ is either $0$ or $2$, and if there is no cap on short selling, then 
an investor can purchase this information for a modest cost to make infinite 
profit, leading to an arbitrage because starting with zero initial wealth $W_0=0$ 
one can construct a portfolio such that ${\mathbb E}[W_1]=\infty$ under any 
measure equivalent to ${\mathbb P}$. In contrast, in our theory we would like 
the cost of information be infinite, if this information were to provide infinite benefit. 

With this in mind, we deduce that the `correct' pricing formula is given by solving 
\begin{eqnarray}
{\mathbb E}_\pi[U(W_p^*)] = {\mathbb E}_\pi[U(W_{\pi,c}^*)] \label{eq:17} 
\end{eqnarray}
for $c$. In other words, we consider with hindsight what would have happened 
to the uninformed strategy, and compared this with the informed strategy. In this 
way, a consistent price for information can be deduced. We propose this to be the 
basis with which information can be priced using a utility indifference argument. 

\begin{figure}[t]%[!h]
\center
\includegraphics[width=1\textwidth,clip]{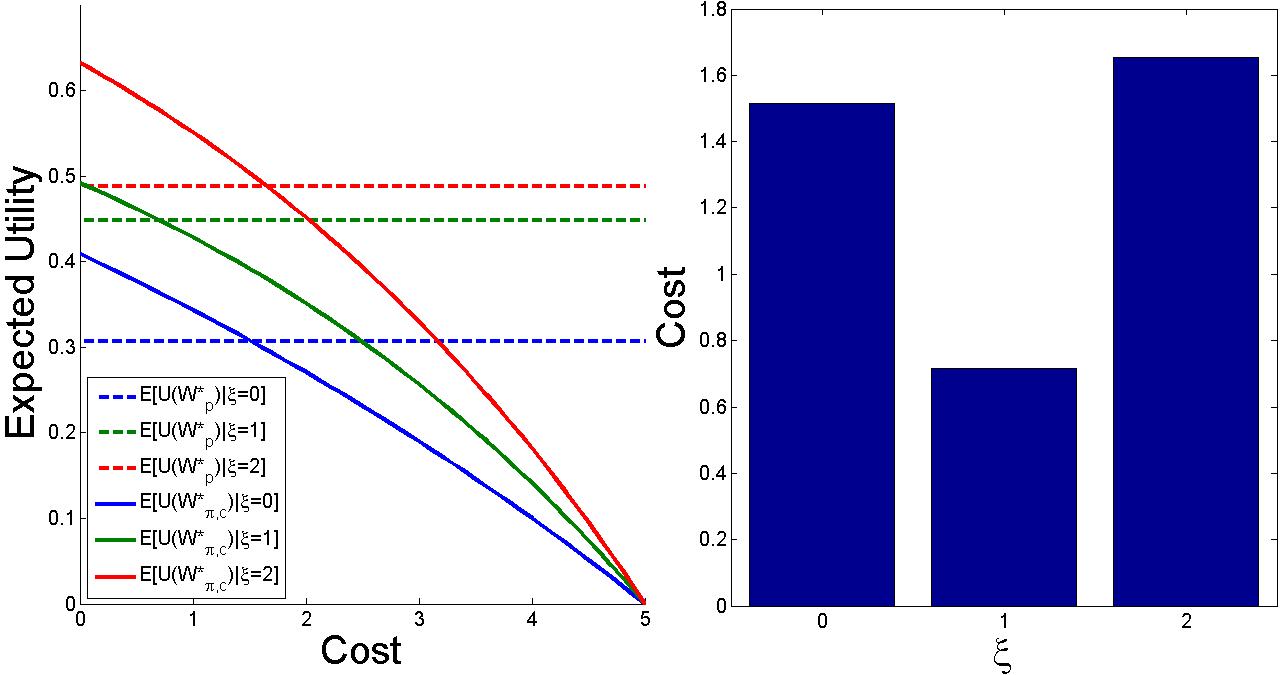}
 \caption{Left: Expected utilities ${\mathbb E}_\pi[U(W_p^*)]$ for $\xi=0,1,2$ 
 as a function of the cost $c$ are shown here as three horizontal lines. For each 
 value of $\xi$ we also plot ${\mathbb E}_\pi[U(W_{\pi,c}^*)]$, when short selling 
 is strictly prohibited. When $c$ is 
 sufficiently small, the latter utilities are higher, but as the cost increases, they 
 decrease. At $c=c^*$ the two utility functions for each $\xi$ agree and determine 
 the unique equilibrium price for which the investor is indifferent. Here we have 
 chosen parameters to be $\alpha=0.1$, $s_0=0.5$, $W_0=5$, $p=0.4$, $q=0.7$, 
 and $\delta=0.95$. The 
 numerical values for the cost for each $\xi$ are plotted as a histogram on the 
 right side. In this example, we have $c^*(0)=1.51$, $c^*(1)=0.71$, and 
 $c^*(2)=1.65$. The average cost is ${\bar c}^*=1.10$. 
\label{fig:1}}
\end{figure} 

In the case of the above example, (\ref{eq:17}) amounts to solving 
\begin{eqnarray}
\re^{-\alpha \delta^{-1} (W_0-\phi^* s_0)} \left(\pi_\xi + (1-\pi_\xi) 
\re^{-\alpha \phi^*} \right) = 
\re^{-\alpha \delta^{-1} (W_0-c-\varphi^* s_0)} 
\left(\pi_\xi + (1-\pi_\xi)\re^{-\alpha \varphi^*} \right)
\end{eqnarray}
for $c$. This gives 
\begin{eqnarray} 
c^*= \frac{\delta}{\alpha} \ln \left[ 
\frac{\pi_\xi \re^{\alpha {\bar s}_0 \phi^*}+
(1-\pi_\xi)\re^{\alpha\phi^*({\bar s}_0
-1)}}{ \pi_\xi \re^{\alpha {\bar s}_0 \varphi^*}+
(1-\pi_\xi)\re^{\alpha\varphi^*({\bar s}_0-1)}}  
\right], 
\end{eqnarray} 
which makes it apparent that when $\phi^*=\varphi^*$, i.e. when $\xi$ provides no 
additional information (which can happen in the present example if for instance 
$p=q=\frac{1}{2}$ and $\xi=1$), or when the additional information provided by $\xi$ 
results in no change of the strategy, the cost is identically zero. Note that the 
dependence of $\pi_\xi$ on $\xi$ implies that the value of $c^*$ also depends on 
$\xi$. Thus, for instance if $\xi=0$ or $\xi=2$, and if unlimited short selling is 
allowed, then $c^*=\infty$ as desired. In many cases the dependence on $\xi$ is 
natural, because different values of $\xi$ embody different information contents. 
On the other hand, there are likewise many cases where it is desirable to associate 
fixed price for information. This applies to the provision of generic information, 
whereas for the purchase of a specialised information, it might be unreasonable 
to expect a flat rate for the information, irrespective of its content or quality. We shall therefore consider both cases. 

If short selling is prohibited, then even with the knowledge $\xi=0$ or $\xi=2$ an 
investor can only make finite profit. Hence in this case the cost should also be 
strictly bounded. In figure~\ref{fig:1} on the left panel we plot 
${\mathbb E}_\pi[U(W_p^*)]$ for three values of $\xi$ as functions of $c$ (clearly, 
${\mathbb E}_\pi[U(W_p^*)]$ for each value of $\xi$ is constant in $c$), and 
compared these with ${\mathbb E}_\pi[U(W_{\pi,c}^*)]$ as functions of $c$, when 
short selling is forbidden. For each value of $\xi$ 
the intersection of the two functions determine the cost $c^*(\xi)$, shown as a 
histogram also in figure~\ref{fig:1}. For the parameter values chosen in this 
example, the `upside' information (when $\xi=2$) is worth a little more than the 
`downside' information (when $\xi=0$). In this way we are able to assign prices 
for information of specific quality. If we are interested in assigning a flat rate 
${\bar c}^*$, then we average individual cost according to 
\begin{eqnarray}
{\bar c}^* = \sum_{k} {\mathbb P}(\xi=\xi_k) c^*(\xi_k) . 
\label{eq:z21}
\end{eqnarray}
In the present example we have ${\mathbb P}(\xi=0)=pq$, ${\mathbb P}(\xi=1)=
p(1-q)+(1-p)q$, and ${\mathbb P}(\xi=2)=(1-p)(1-q)$, which can be used to 
determine ${\bar c}^*$. Since the \textit{a priori} expectation of our pricing 
formula (\ref{eq:17}) gives the Grossman-Stiglitz formula (\ref{eq:17.0}), we see 
that solutions to (\ref{eq:17.0}) and (\ref{eq:17}) correspond, respectively, to an 
annealed average and a quenched average often considered in the spin-glass 
theory. In particular, on account of Jensen's inequality we see that the 
Grossman-Stiglitz cost is the lower bound of our flat-rate cost ${\bar c}^*$. 

Having introduced our framework for information pricing, the next objective in this 
paper is to express the cost in terms of the quantity of information so that we 
are able to quote the price in the form, for example: ``$\pounds$X for Y bits of 
upside information''. This is highly desirable because the representation of the 
cost as a function of $\xi$ is not very practical. Our definition of information as 
a quantity that generates the transformation (\ref{eq:3}) allows us to proceed by 
the consideration of relative entropy between the \textit{a priori} and the 
\textit{a posteriori} probabilities. The examples considered above, however, are 
too restrictive, because the \textit{a priori} and the \textit{a posteriori} probabilities 
are not absolutely continuous with respect to each other, and thus relative 
entropy cannot be defined. Indeed, the idea that an information provider 
can ascertain the future value $s_1$ at time $0$, which would be the case if 
the sampled value of $\xi$ is either $0$ or $2$, is somewhat artificial and 
unrealistic. We therefore consider another simple example in which this issue 
does not arise. 

\noindent \textit{Example 2.1: uninformed investor}. We assume the setup as in 
Example 1.1 above, except that the random variable $s_1$ is assumed normally 
distributed with mean $\mu$ and variance $\sigma^2$. Thus, the model is similar 
to the Bachelier model, rather than a geometric Gaussian model, in that the asset 
price can take negative values. A straightforward Gaussian integration then gives 
the expected utility of terminal wealth as: 
\begin{eqnarray}
{\mathbb E}_p[U(W_1)] = 1-\exp\left[ -\alpha \left( \delta^{-1}(W_0-\phi s_0 ) 
+\phi \mu - \half \alpha \phi^2 \sigma^2 \right)\right] .
\end{eqnarray}
It follows that the optimal allocation strategy is determined by  
\begin{eqnarray}
\phi^* = \frac{\mu - {\bar s}_0}{\alpha \sigma^2}, \qquad \theta^* = 
W_0- \frac{(\mu - {\bar s}_0)s_0}{\alpha \sigma^2} .
\end{eqnarray}
If there is a restriction on short selling, then these results hold provided that 
$0 \leq \phi^* \leq W_0/s_0$; otherwise, $\phi^*=0$ or $\phi^*=W_0/s_0$.
\hfill$\diamond$

\noindent \textit{Example 2.2: informed investor}. An informed investor is able 
to purchase the sampled value of $\xi$ given by (\ref{eq:1}) 
above, except that the noise variable $\epsilon$ is now assumed normally 
distributed with mean zero and variance $\sigma_\epsilon^2$. In this case, a 
standard result in Gaussian filtering theory \cite{wiener2} shows that the 
\textit{a posteriori} probability law for $s_1$ given $\xi$ is also normally 
distributed: 
\begin{eqnarray}
s_1|_\xi \sim N\left( \frac{\mu \sigma^2_\epsilon+ \xi \sigma^2}
{\sigma^2_\epsilon+ \sigma^2},\sqrt{\frac{\sigma^2  \sigma^2_\epsilon}
{\sigma^2 +\sigma^2_\epsilon}}\right) \equiv N(\mu_\xi,\sigma_\xi)
\end{eqnarray}
The optimisation problem thus reduces to maximising  
\begin{eqnarray}
u_c(W_0) = 1-\exp\left( -\alpha \left( \delta^{-1} (W_0-c-\varphi s_0 ) +\varphi 
\mu_\xi - \half \alpha \varphi^2 \sigma^2_\xi \right)\right). 
\end{eqnarray}
The result is 
\begin{eqnarray}
\varphi^* = \frac{\mu_\xi - {\bar s}_0}{\alpha \sigma^2_\xi}, \qquad 
\vartheta^{*} = W_0-c- \frac{(\mu_\xi - {\bar s}_0)s_0}{\alpha \sigma^2_\xi}. 
\label{eq:25}
\end{eqnarray}
As before, if short selling is not permitted, we have the conditions 
$0 \leq \varphi^* \leq (W_0-c)/s_0$ for which (\ref{eq:25}) is 
valid; otherwise we must take boundary values. \hfill$\diamond$ 

\begin{figure}[t]%[!h]
 \center
 \includegraphics[width=0.6\textwidth,clip]{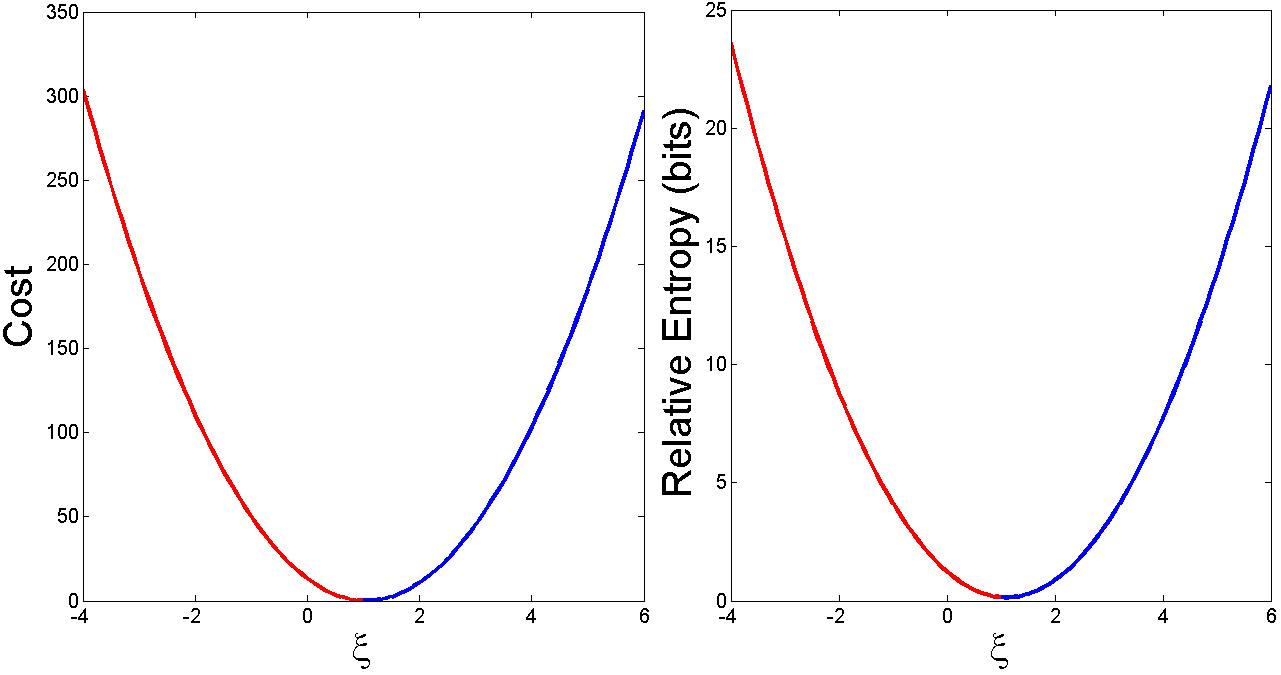}
 \caption{Left: Cost $c^*(\xi)$ as a function of the realised value of $\xi$. Right: 
 Relative entropy $I(\xi)=I(\pi_\xi|p)$ as a function of the realised value of 
 $\xi$. 
 The parameter values used here are: $s_0=1$, $\delta=0.95$, $\alpha=0.1$, 
 $W_0=10$, $\mu=1.1$, $\sigma^2=0.2$, $\mu_\epsilon=0$, and 
 $\sigma^2_\epsilon=0.2$. The red curves correspond to pieces of downside 
 information (i.e. values of $\xi$ that indicate decrease in $s_1$ as compared to 
 the \textit{a priori} expectation) and the blue curves correspond to upside 
 information.}
 \label{fig:2}
\end{figure}

Let us evaluate the cost of information by the principle (\ref{eq:17}). 
This amounts to finding the value for $c$ that solves 
\begin{eqnarray}
&& \exp\left[ -\alpha \left( (W_0-\phi^* s_0 )\delta^{-1} +\phi^* \mu_\xi - 
\half \alpha \phi^{*2} \sigma_\xi^2 \right)\right]  \nonumber \\ && \qquad = 
\exp\left[ -\alpha\left( (W_0-c-\varphi^* s_0 ) \delta^{-1} +\varphi^* \mu_\xi 
- \half \alpha \varphi^{*2} \sigma^2_\xi \right)\right], 
\end{eqnarray}
with the solution 
\begin{eqnarray}
c^* = \half \alpha \delta \sigma_\xi^2 (\phi^*-\varphi^*)^2 . 
\label{eq:27} 
\end{eqnarray}
This provides the cost $c^*(\xi)$ as a function of the sampled value of $\xi$. 
That is, before the purchase is made, a client of the information provider can 
agree to the price structure (\ref{eq:27}) so that depending on which value of 
$\xi$ the information provider produces, the client will pay the cost appropriate 
for that information. If the use of a flat rate ${\bar c}^*$ for the cost is desirable, 
then we average $c^*(\xi)$ over $\xi$ under the suitable Gaussian measure. 
This is given by 
\begin{eqnarray}
{\bar c}^* = \frac{1}{2\alpha} \delta \sigma^2_\xi \left[ (\mu -\bar{s}_0)^2 
\left(\sigma^{-2} - \sigma_\xi^{-2}\right)^2 + \frac{\sigma^2}
{\sigma^2_\xi \sigma^2_\epsilon} \right] . 
\end{eqnarray}

\begin{figure}[t]%[!h]
 \center
 \includegraphics[width=1\textwidth,clip]{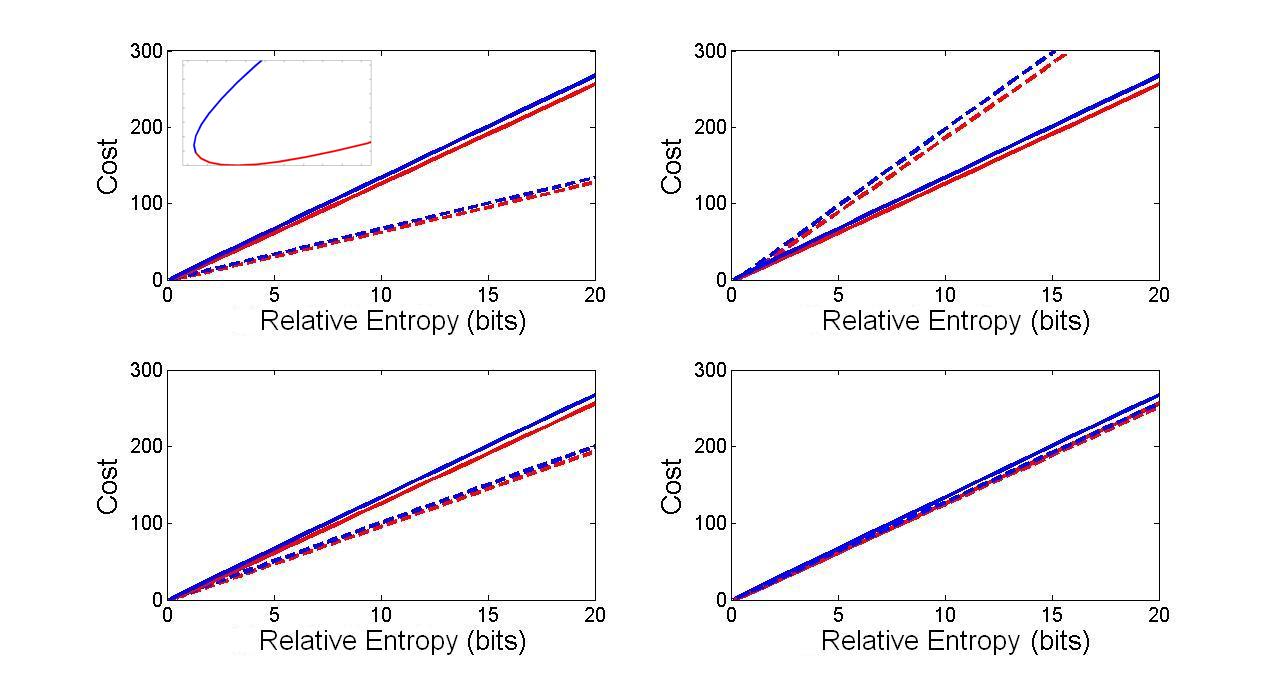}
 \caption{The cost $c^*(I)$ as a function of the quantity of information $I$, when 
 there is no constraint for short selling. The solid lines in all four plots (they are 
 identical) correspond to the parameter values used in figure~\ref{fig:2}. 
 While 
 keeping other parameters fixed, the following changes in parameter values 
 are used, represented by dashed lines in the plots. Top left: $\alpha$ has been 
 increased to $\alpha=0.2$; top right: $\sigma^2$ has been increased to 
 $\sigma^2=0.4$; bottom left: $\sigma^2_\epsilon$ has been increased to 
 $\sigma^2_\epsilon =0.4$; and bottom right: $\delta$ has been decreased to 
 $\delta=0.92$. The red curves correspond to downside information, 
 and the blue curves correspond to upside information. 
 The insert shows the behaviour of $c^*(I)$ close to the origin where $c^*(I)$ 
 is decreasing in $I$.}
\label{fig:3}
\end{figure}

As indicated above, we wish to associate cost with the quantity of 
information. For this purpose, we consider the relative entropy: 
\begin{eqnarray}
I(\pi_\xi|p ) &=& \int_{-\infty}^\infty 
\pi_\xi(s) \ln\left(\frac{\pi_\xi(s)}{p(s)} \right) \rd s \nonumber \\
&=& \half \left[ \ln\left(\frac{\sigma^2}{\sigma^2_\xi} \right) +  
\frac{\sigma^2_\xi}{\sigma^2}-1   \right]+ \half \frac{(\mu_\xi-\mu)^2}{ \sigma^2} 
\label{eq:28}
\end{eqnarray}
between the prior and the posterior densities. It should be evident from the 
decomposition $\xi=s_1+\epsilon$ that there is a critical value $\xi^\dagger$, 
given by the \textit{a prior} expectation of $s_1$, such that the \textit{a posteriori} 
density is least informative, and such that the information content increases as 
the value of $\xi$ increases or decreases away from this critical level. In other 
words, relative entropy is monotonically decreasing in $\xi$ if $\xi<\xi^\dagger$, 
and monotonically increasing in $\xi$ if $\xi\geq \xi^\dagger$. One might therefore 
expect that the cost should also exhibit the same monotonic dependence on 
$|\xi-\xi^\dagger|$. Surprisingly, however, this is not the case. The pricing formula 
(\ref{eq:27}) in the Gaussian model shows that if the additional information does 
not alter portfolio positions, then an investor is not going to pay for that 
information. As a consequence, minimum information can be more costly (i.e. 
useful) than a larger quantity of information. 

\begin{figure}[t]%[!h]
 \center
 \includegraphics[width=0.6\textwidth,clip]{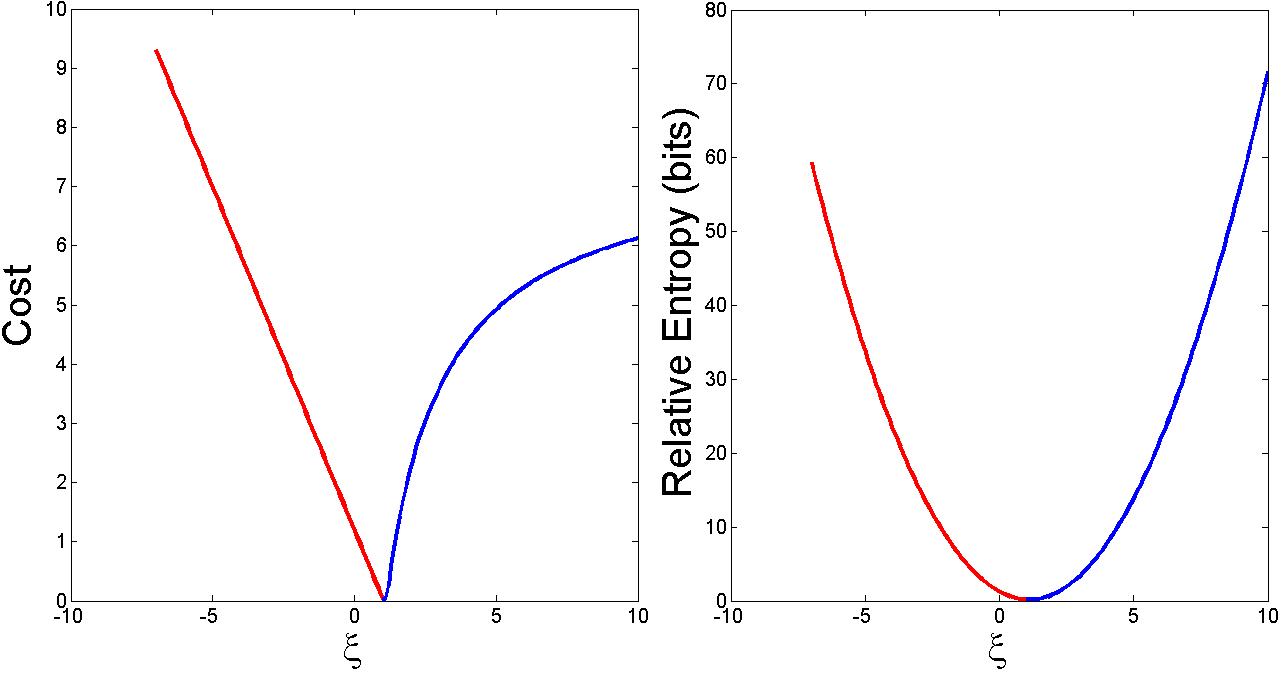}
 \caption{
 Left: Cost $c^*(\xi)$ as a function of the realised value of $\xi$ when short 
 selling is strictly prohibited. Right: Relative entropy $I(\xi)=I(\pi_\xi|p)$ as a 
 function of the realised value of $\xi$. The cost $c^*(\xi)$ measures the 
 usefulness of information, while relative entropy $I(\xi)$ measures the quantity 
 of information. The parameter values used here are: $s_0=1$, $\delta=0.95$, 
 $\alpha=0.1$, $W_0=10$, $\mu=1.1$, $\sigma^2=0.2$, $\mu_\epsilon=0$, and 
 $\sigma^2_\epsilon=0.2$. The red curves correspond to pieces of downside 
 information (i.e. values of $\xi$ that indicate decrease in $s_1$ as compared to 
 the \textit{a priori} expectation) and the blue curves correspond to upside 
 information.
 }
 \label{fig:4}
 \end{figure}
 
In figure~\ref{fig:2} we plot the cost function $c^*(\xi)$ where there are no 
constraints for short selling, and the relative entropy $I(\xi)$. By inverting these 
two functions we can determine the cost $c^*(I)$ as a function of the information 
content, as shown in figure~\ref{fig:3} for a range of parameter values. Note that 
in the numerical plots we have converted the logarithmic basis so that relative 
entropy $I$ is measured in the conventional binary units of `bits' rather than 
the Shannon units used in (\ref{eq:28}). Hence the horizontal axis represents 
information content expressed in binary units.  
We observe that costs of upside information in these examples are 
always slightly higher than those of downside information, and that in each case 
the cost is approximately a linear function of the quantity of information. We also 
find that the cost in the unconstrained case is a decreasing function of the risk 
aversion parameter $\alpha$ in the utility function, an increasing function of the 
signal uncertainty $\sigma$, a decreasing function of the noise uncertainty 
$\sigma_\epsilon$, and a slightly increasing function of the discount factor 
$\delta$. 

The cost structure changes somewhat if we restrict short selling. This is because 
the upside gain from exploiting additional information is restricted in this case. In 
particular, if short selling (including cash borrowing) is prohibited, then the cost is 
bounded above by the initial wealth $W_0$. We plot in figure~\ref{fig:4} the cost 
$c^*(\xi)$ as a function of $\xi$ in the case for which short selling is strictly 
forbidden. The result reveals some interesting insights: a client of an information 
provider is willing to pay significantly higher costs for large downside information, 
as compared to upside information, and the increase in $c^*(\xi)$ for upside 
information is satiated relatively early. This feature of course is closely related 
to the model setup chosen here: Since $s_1$ is assumed Gaussian, purchase 
of the stock by any amount can result in an unbounded loss. Therefore, investors 
are willing to spend the totality $W_0$ of the initial budget as a protection against 
large losses, leaving behind no asset for any investment. Conversely, investors 
are inclined not to spend significant portions of the initial budget on upside 
information, because if there are opportunities to profit from investments, they 
would not want to lose these opportunities by spending all the budgets. 

\begin{figure}[t]
 \center
 \includegraphics[width=1\textwidth,clip]{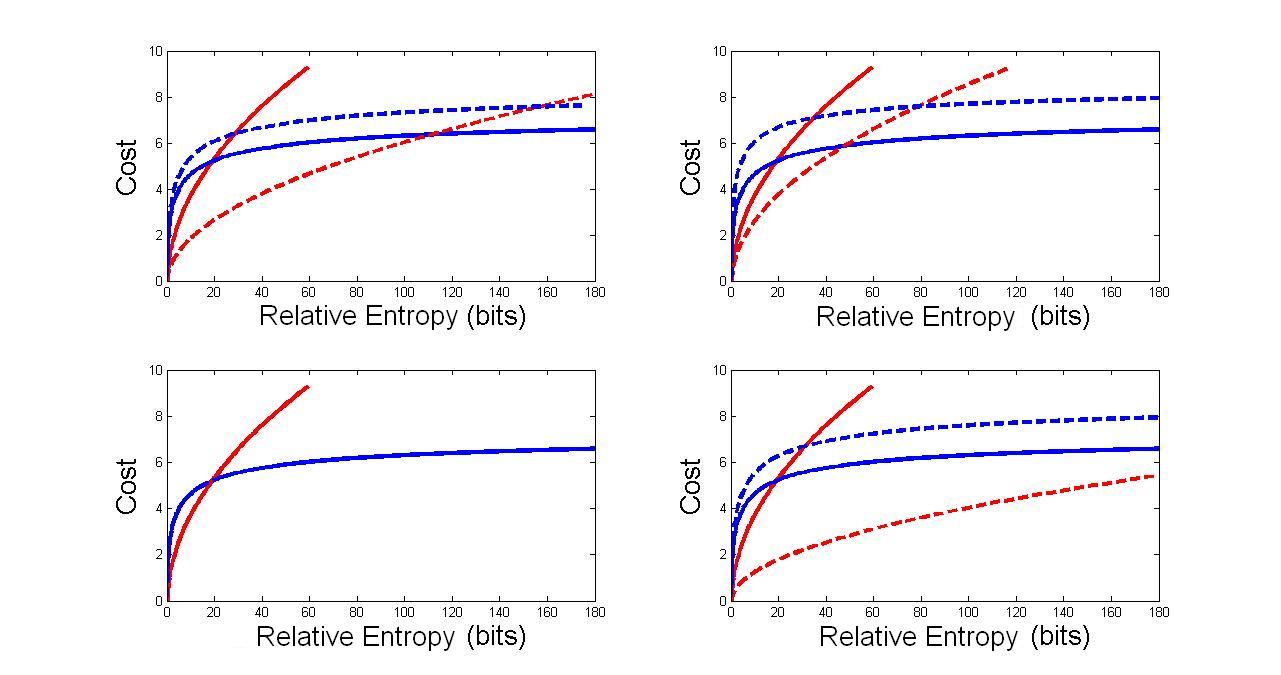}
 %{Plot9CostvsREGaussianCont.jpg}
 \caption{
 The cost $c^*(I)$ as a function of the quantity of information $I$, when 
 short selling is strictly prohibited. The solid lines in all four plots (they are 
 identical) correspond to the parameter values used in figure~\ref{fig:4}. While 
 keeping other parameters fixed, the following changes in parameter values 
 are used, represented by dashed lines in the plots. Top left: $\alpha$ has been 
 increased to $\alpha=0.2$; top right: $\sigma^2$ has been increased to 
 $\sigma^2=0.4$; bottom left: $\sigma^2_\epsilon$ has been increased to 
 $\sigma^2_\epsilon =0.4$; and bottom right: $\delta$ has been decreased to 
 $\delta=0.92$. The red curves correspond to downside information, 
 and the blue curves correspond to upside information.
 }
\label{fig:5}
 \end{figure}

Although the cost structure changes when there are constraints, we can again 
invert these relations to quote unique prices for upside 
and downside information. The results are shown in figure~\ref{fig:5} for a range 
of parameter values. We observe that when there are constraints the price of 
information exhibits somewhat nontrivial behaviours as we change the model 
parameters. 

\begin{figure}[t]
\center
 \includegraphics[width=1\textwidth,clip]{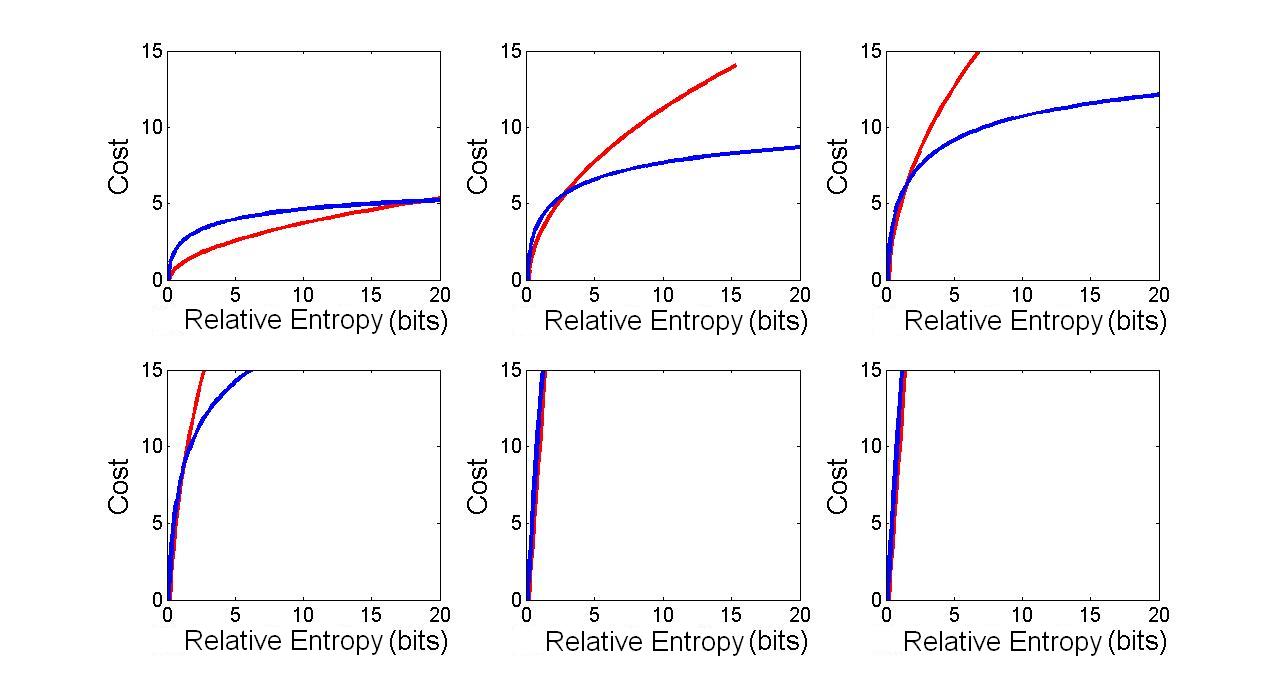}
 \caption{Interpolation of the information price $c^*(I)$ from the strictly constrained 
 case to the unconstrained case. Top row, from left: (i) borrowing is strictly forbidden; 
 (ii) borrowing is capped at $W_0/2$; (iii) borrowing is capped at $W_0$. Bottom row, 
 from left: (i) borrowing is capped at $2W_0$; (ii) borrowing is capped at $5W_0$; 
 (iii) unlimited borrowing permitted.}
\label{fig:6}
\end{figure}

The two cases examined here can be interpolated by gradually relaxing the 
borrowing cap, as shown in figure~\ref{fig:6}. These intermediate cases are 
often more realistic because of leveraging. In the case of the example considered 
here we observe that by raising the cap of borrowing to five times the total 
wealth the net effect is already equivalent to allowing for unlimited borrowing. 
The effect of leveraging is also of interest in showing in which way the upside 
and downside information costs intertwine. 

We conclude by remarking that the existence of information asymmetries is 
not only fundamental to modern microeconomic theory \cite{NobelIntro} but 
also constitutes the basis for the existence of information providers, whose 
roles are becoming increasingly more important in modern society. Our main 
objective here is to demonstrate in which way valuable information can be 
priced in a rational manner. Although the models used here to illustrate our 
pricing theory are simple, they nevertheless provide a useful conceptual guideline 
for evaluating information in a variety of contexts. It would be of interest to 
extend the single-period models considered here to continuous-time models 
within the context of utility indifference pricing theory (as outlined, e.g., in 
\cite{Schweizer,Imkeller,Monoyios08}).

\vskip 10pt 

\noindent The authors thank Bernhard Meister for comments. DCB 
acknowledges Fraunhofer ITWM, Kaiserslautern, for hospitality while 
part of this work was carried out.

%\newpage

\end{document}